# Identification of Desired Pixels in an Image Using Grover's Quantum Search Algorithm

**Basit Iqbal[1], Harkirat Singh[1]**

**Abstract**
Quantum Information Theory promises to speed up computation so is observed in real quantum computers as proved to its classical counterpart. This revolutionizes every field linked directly or indirectly with computation. Grover algorithm in quantum information gives quadratic speed up in unstructured database search. With the availability of public online resources for quantum computers like by IBM, quantum image processing came into the picture for making use of quantum computers in the image processing field.

Our research interest is to find all darker pixels in a 2x2 grayscale image using Grover's algorithm. We studied it by two different ways. In the first method, I ran Grover's algorithm on the python generated classical image. In the second method, I converted a python generated 2x2 image into a quantum image and then ran the Grover's algorithm to locate the darker pixels. As has been observed in complexity analysis, Grover's unstructured search has the $O(2^n)$ while as for classical schemes $O(2^{2n+2m})$, where m and n denote the dimensions of the image.

✉@ [1]Basit Iqbal

Department of physics, National Institute of Technology Srinagar, Hazratbal, Srinagar, J&K-190006
khanbasitiqbal@gmail.com

[1]Harkirat Singh

Department of physics, National Institute of Technology Srinagar, Hazratbal, Srinagar, J&K-190006
harkirat@nitsri.ac.in

# 1 Introduction

Quantum Information Theory promised speed up computation in quantum computers as compared to its classical counterpart. This is revolutionizing every field linked directly or indirectly with computation. With the availability of public online resources for quantum computation like by IBM [23] quantum image processing can be realized by making use of quantum computation in the image processing discipline. The discipline of quantum computation combined with image processing and with the help of principles of quantum mechanics like coherence, superposition and entanglement, helps us to overcome the limits of efficiency of classical image processing. Also, it helps to take the benefits of quadratic/exponential speed up of quantum algorithms in classical image processing.

Quantum computers take pixel locations and their attuned shades of a classical image, and present the integrated information as a quantum state [1-4]. Representation of quantum images has been studied extensively, in last few years. The various quantum representations of images are: Qubit Lattice [5], Entangled Image [6], Real Ket [7], Flexible Representation of Quantum Images (FRQI) [8], Multichannel Representations of Quantum Images (MCRQI) [9], Normal Arbitrary Quantum superposition State (NASS) [10], Quantum Representation of Log polar Images (QUALPI) [11], and Novel Enhanced Quantum Image Representation (NEQR) [12]. The proven efficiency of speed up using quantum algorithms has been studied in [14-16] While classical computation takes O(N) evaluations in an unsorted database search, where N is the of items in the list, Grover's algorithm giving quadratic speedup, asymptotically optimal [17], by doing the same job in $O(N^{1/2})$ evaluations. Grover's algorithm takes the search item and submits it to the oracle which flips the sign if the item being searched and passes it to the diffuser operator. Diffuser operator amplifies the probability amplitude of the item being searched and decreases the probability amplitude of the other items in the search list [18, 19].

One of the interesting topics of quantum image processing is hiding of digital information with the help of quantum watermarking for protecting and sending any digital information. No cloning theorem and Heisenberg's uncertainty principle are principles of quantum mechanics which help us to hide the data in an image. Now, as the field of quantum image processing is developing new steganalysis techniques are being developed to hide the data in a quantum image. In Quantum image we hide the data in such a manner that it is not distinguishable to any except the receiver to decode the secret information in it [20].

NEQR is a modified FRQI model which uses the tensor product of two qubit sequences to accumulate the information of whole image [12]. First qubit sequence denotes the grayscale measure of the pixel and second sequence denotes corresponding pixel position in image [12]

As the field of quantum computing is grooving the curiosity for their applications in various fields linked with computing is also increasing to get the benefits of their high efficiency like speed up. So is for the field of stenography. We worked on the methods for identification of darker pixels using Grover's algorithm to find the application of the method in classical and quantum stenography. In first experiment, we generated an image, obtained data from the image, defined a threshold for darker pixels and applied Grover's algorithm to find the darker pixels in the image. In second experiment first of all we generated a 2x2 image using an arbitrary array; secondly, convert that image into a NEQR quantum image; and finally, turned to our main focus, used the Grover's algorithm to find all the darker pixels in the image below certain threshold.

## 2 Computational Details

Computational details of our experiment fall in two categories for two experiments. One experiment is done in the framework of Cirq and the second frame work is of Qiskit which are open sources for doing Quantum Computing. [23]

### 2.1 Experiment in the Cirq framework:

Cirq is an open-source Noisy Intermediate Scale Quantum Computer framework provided by Google for dealing with quantum circuits such as writing, manipulating and optimizing them, and can also run them on their quantum computers and quantum simulators. In this at first, we generated a 2x2 image from an array using Python Image Library (PIL) and saved it as a grayscale image (fig.1). Secondly, we opened the saved image in python and obtained the data i.e., pixel intensities, as an array. In python programming, intensity of images varies from 0-255, with 0 denoting complete black and 255 denoting complete white pixel. Thirdly, we defined a threshold of 100 i.e., all the pixels with intensities less than 100(i.e., 0-100) are considered as dark shade pixels and the pixels with intensities greater than 100(i.e., 100-255) are considered as white shade. Our aim is to locate these dark shade pixels.

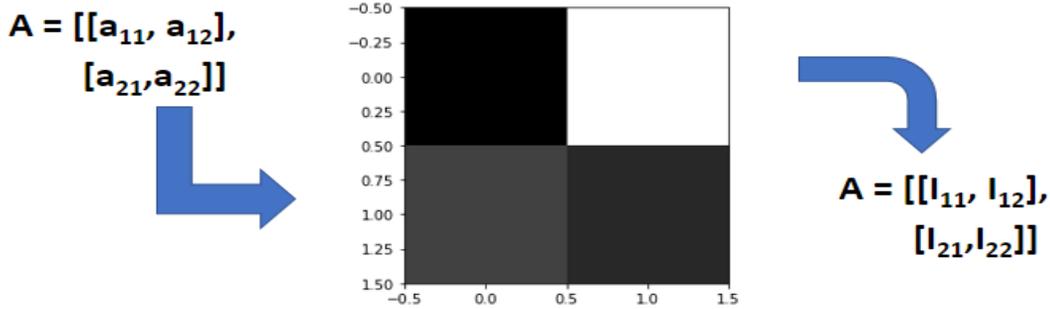

**Fig. 1.** Generating an image from an array and converting that image into an array

Finally, Grover's algorithm is iterated on all pixels, one by one. Grover's algorithm takes these darker pixels in the unsorted search, amplifies the amplitude and gives the coordinates of all the darker pixels.

## 2.2 Experiment in the Qiskit framework:

Qiskit is another opensource framework provided by IBM which use Python Software Library for writing and manipulating quantum programs. One can run these programs on IBM Quantum Experience devices or on IBM quantum Simulator.

NEQR image representation composed of two parts. First part denotes the grayscale measure of the pixel. For gray scale image, the intensity value varies from 0 to 255 and in binary numbers we require 8 bits to represent an intensity value. Second part represents the pixel by their position [12]. Further, column and row numbers represent x-values and y-values respectively. In general, a $2^n \times 2^n$ resolution image can be written in NEQR representation as a quantum state $|I\rangle$ [21] as

$$|I\rangle = \frac{1}{2^n} \sum_{Y=0}^{2^{2n-1}} \sum_{X=0}^{2^{2n-1}} |f(X,Y)\rangle |YX\rangle = \sum_{Y=0}^{2^{2n-1}} \sum_{X=0}^{2^{2n-1}} \left| \bigotimes_{i=0}^{q-1} \right\rangle |C_{YX}^i\rangle |YX\rangle$$

So, for the 2x2 pixel image in fig. 1 with pixel intensities 0, 255, 65 and 40 NEQR state will be represented as

$$|I\rangle = (1/\sqrt{2}) (|0\rangle|00\rangle + |255\rangle|01\rangle + |65\rangle|10\rangle + |40\rangle|11\rangle) =$$

$$\frac{1}{\sqrt{2}}(|00000000\rangle|00\rangle + |11111111\rangle|01\rangle + |01000001\rangle|10\rangle + |00101000\rangle|11\rangle)$$

In this experiment, firstly, we generated a 2x2 grayscale image was generated from an array and saved it, as in first experiment. Secondly, that image is converted into an NEQR (Novel Enhanced Quantum Image Representation) image representation using the code provided by IBM Qiskit Textbook [21]. Again, the threshold is of hundred i.e., all the pixels with values less than 100 (i.e., 0-100) are considered as dark shade pixels and the pixels with higher values (i.e., 100-255) are considered as light shade pixels. Our aim is to locate all the pixels with intensity less then threshold of 100. Now, Grover's algorithm is run to amplify the states of the pixels with intensity less than 100.

Contrary to first experiment, in the second experiment, we converted an image (classical) into a quantum state i.e., NEQR image representation and then applied Grover's algorithm. Further in second experiment, all the states representing black pixels are amplified in single run, not by iterating one by one.

## 2 Results and Discussions

The simulations lie on the basis of matrix theory with Grover's algorithm and modified according to our experiment (i.e., some functions and commands are added). Both of the two experiments are done using python in Jupyter Notebook. The experiments are done on images of 2x2 pixel size. The images and the corresponding results with Google Cirq [22] and IBM Qiskit [21] are discussed below:

**3.1 Results of experiments with Google Cirq:**

This study is done using two qubit Grover's Algorithm [22]. In this investigation, we chose pixel intensities of 30, 60 80 and 200 with different pixel positions. Since the threshold is of 100-pixel intensity, so as expected, three darker pixels are identified using Grover's algorithm. The circuit diagram of two qubit Grover's algorithm is as shown in fig. 2. in fig. 2 the X, H and M gates are representing the NOT, Hadamard and Measurement gates, respectively. The @-@-X and the @-X gates are representing the CCNOT (Toffoli) and CNOT gates, respectively. The Ancilla is the ancillary qubit just giving the essential support without disturbing the results.

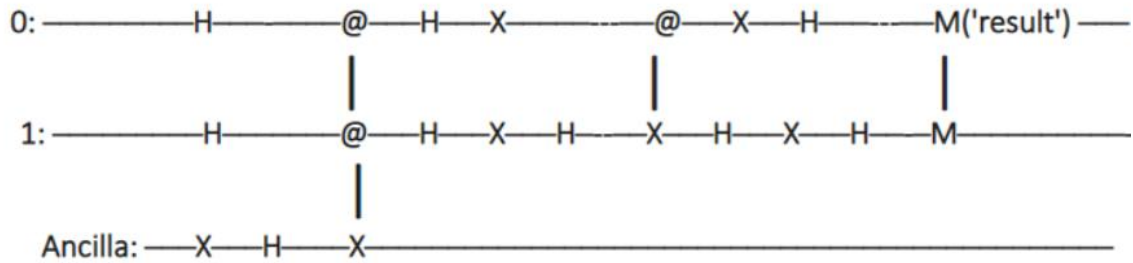

**Fig. 2:** Circuit for Grover's algorithm. @ symbol represents the control of the gates.

We generated four images with the pixel intensities (30, 60, 80 and 200) not altered, only the pixel positions are changed and obtained the results on these four images (3.1(a), 3.2(a), 3.3(a) and 3.4(a)). These results are taken by running the algorithm on Google Cirq's simulator. In the generation of the images, it was taken care of that at least three pixels of the 2x2 pixel image should be with pixel intensity less then threshold of 100. So, there are three pixels which should be located by the Grover's algorithm. As expected, the algorithm maximized the probability amplitude of the states representing the pixels with intensity less than the threshold (100). The algorithm located them and the results as 3d histograms are shown in figure 3 (3.1(b), 3.2(b), 3.3(b) and 3.4(b)).

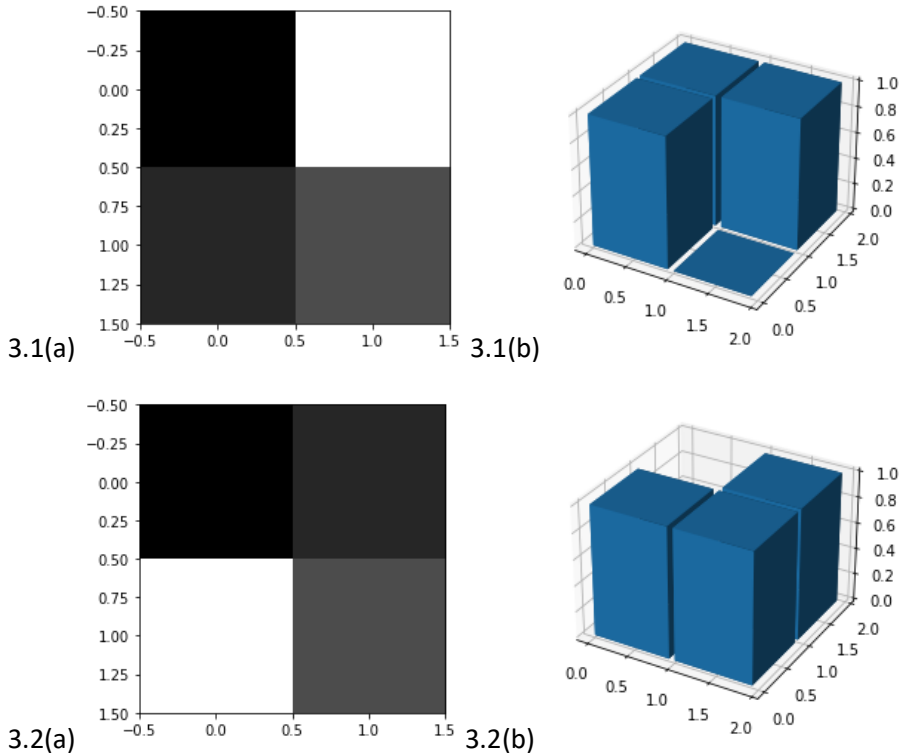

3.1(a)    3.1(b)

3.2(a)    3.2(b)

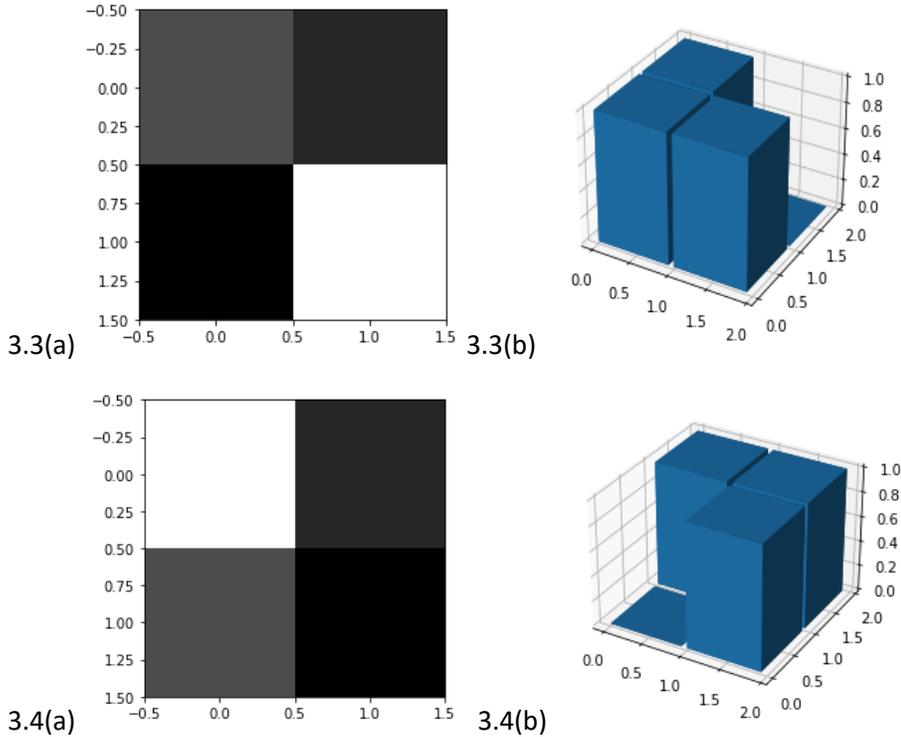

**Fig. 3:** In each row there are 5 images. Each image marked as (a) is the 2x2 pixel image generated and the image, marked as (b) shows the pixels position of the darker pixels in 3d histograms.

## 3.2 Results of experiment in IBM Qiskit

In this study first we converted the python generated 2x2 pixel image into NEQR state by using the code provided by IBM Qiskit online textbook [21] which is modified as per our requirements. Since the image used was a grayscale image, the number of qubits used to show the pixel intensity is eight and the number of qubits sufficient to show position of a pixel in a 2x2 image is two. Thus, ten qubits are required in total. The code is then run-on IBM's quantum simulator. Fig. 4(a) shows the image generated for the purpose which is a 2x2 pixel image with intensities of 0, 20, 40 and 255, fig. 5(c) show the circuit for the conversion of image into a NEQR state and fig. 5(b) show the histogram for the NEQR state generated. In fig. 4 Intensity1, intensity2, …, intensity7 represent the registers to encode pixel intensities. The idx1 and idx2 represent the registers to represent the registers to encode the pixel positions. CCNOT gates are the operators, as proposed by NEQR method, to encode the pixel intensities. If the pixel value is 0 then identity is applied. Python program converts decimal pixel value into binary value. The controls set on the idx1 and idx2 and NOT is applied on the register representing the binary pixel value 1. for example, if the pixel intensity is 100, in binary represented as 01100100 then the NOT will be applied on the intensity2(3[rd]), intensity5(6[th]) and intensity6(7[th]) registers because python reads it in the reverse order as 0, 0, 1, 0, 0, 1, 1 and 0. After encoding the pixel intensities the measurement operators applied.

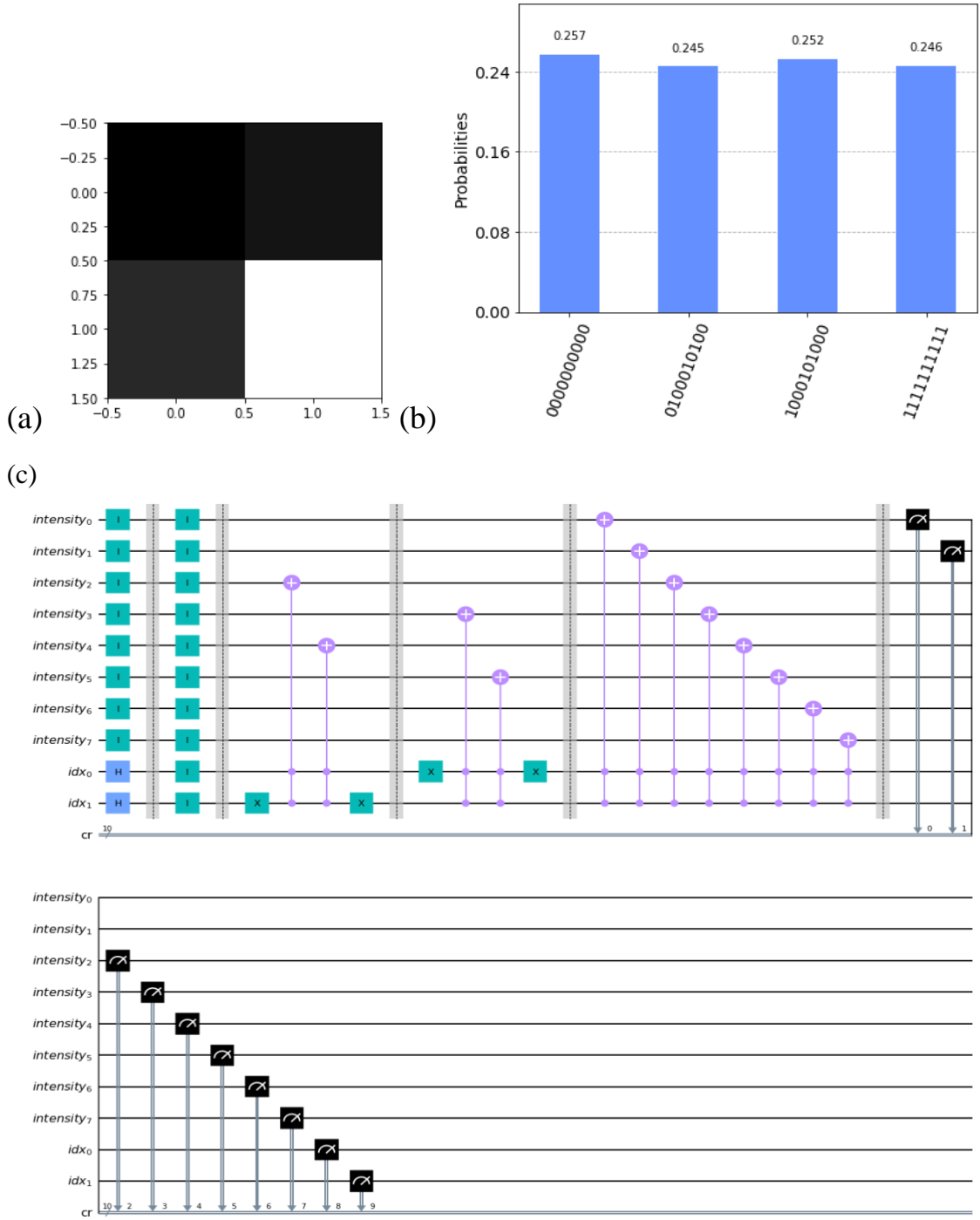

**Fig. 4:** (a) Image (classical) that was used for the conversion to NEQR state (Quantum Image), (b) NEQR histogram for the image, (c) the circuit for the conversion of the image into a NEQR state.

Afterwards, the states were sent to 10 qubit Grover's algorithm for locating all the darker pixels in the image. The circuit of Grover's algorithm for locating the darker pixels can be seen in fig. 5 (a) and the results as can be seen in fig. 5 (b). fig (a) shows the circuit of the Grover's algorithm searching for the pixels searching the states representing the darker pixels. As shown in gig 5(a) fourteen iterations are done by Grover's algorithm to locate the required states. Fig 5(b) shows the result, that is, the three states located by Grover's algorithm that are
(1) the pixel at 00 with pixel intensity 00000000 (in binary), that is, 0 (in decimals)
(2) the pixel at 10 with pixel intensity 00010100 (in binary), that is, 40 (in decimals)
(3) the pixel at 01 with pixel intensity 00101000 (in binary), that is, 20 (in decimals)

(a)

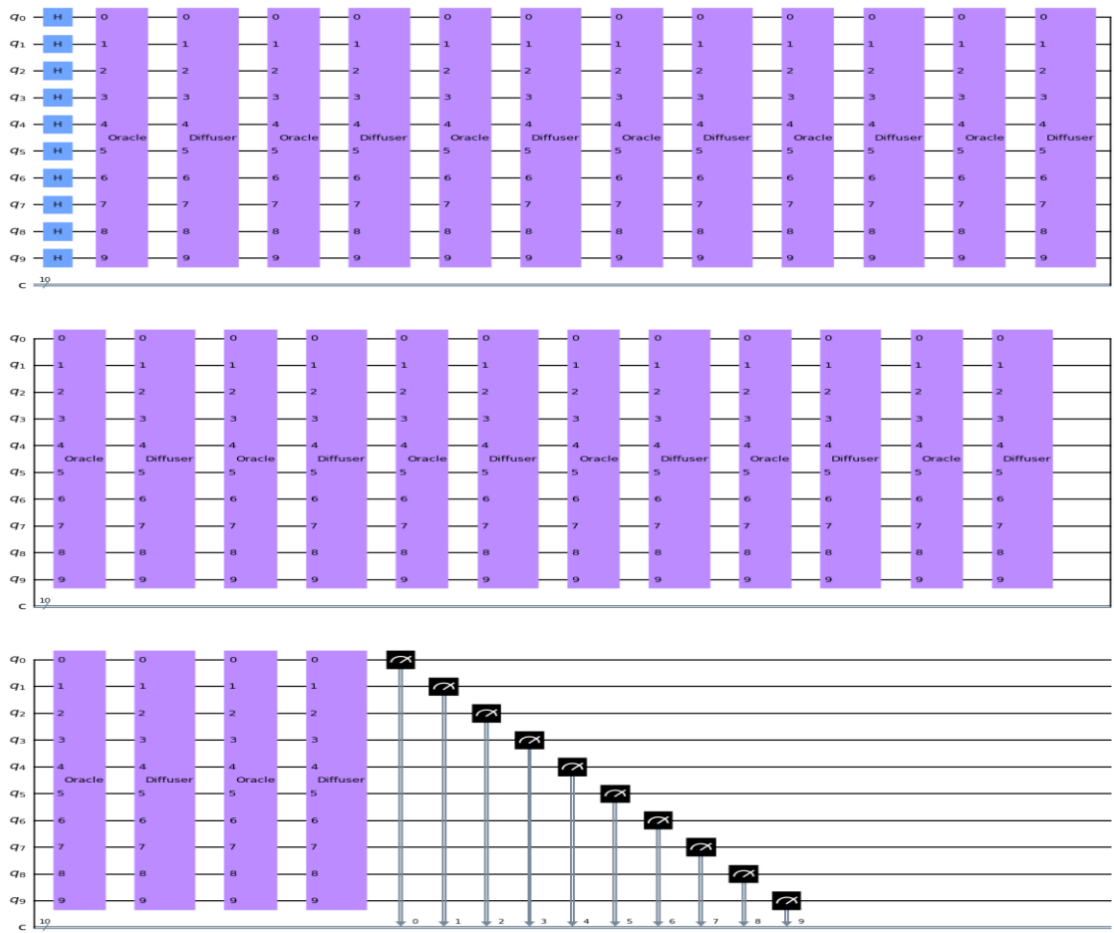

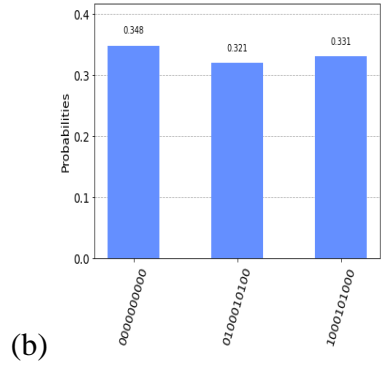

(b)

**Fig. 5:** (a) circuit for 10 qubit Grover's algorithm for finding darker pixel states and (b) the states located by Grover's algorithm

## 3 Conclusion

Identification of darker pixels using Grover's algorithm is studied to find the application of this method in Quantum steganography. In the study done in the Cirq environment four 2x2 pixel images with intensities 0, 40, 60 and 200 at four different locations. Grover's algorithm located all the three darker pixels in the image with intensities less than threshold of 100. In the second study done in qiskit environment, 2x2 pixel image into NEQR quantum state. The Grover's algorithm run to locate all the three darker pixels in the image located the three states representing darker pixels. The experiment done using Cirq environment was semi classical as the conversion of classical to quantum image was not included in this experiment. This method will be useful in stenography in the way for hiding data in classical images as the application of quantum algorithms on classical images has been found useful in terms of quantum benefits over classical like speed up.

In future, the studies will be carried out on larger and colored images. For such purposes we will also need to include Quantum image compression. Further, the use of Quantum error correcting codes will be added to strengthen the practical application of the method.

But the real flavor of quantum benefits comes when we apply quantum algorithms on quantum images. That will further increase our computational benefit in terms of speed. This we have studied in experiment with Qiskit. In this experiment we converted the python generated classical image to a NEQR quantum state and then applied a quantum algorithm and then applied the quantum algorithm, that is, on the state to find all (that is, three) the darker pixels in the image. The results of which are discussed in the Results and Discussion section.